\documentclass[aps,prl,showpacs]{revtex4-1}

\usepackage{amsmath,amssymb}

\newcommand{\Romannumeral}[1]{\uppercase\expandafter{\romannumeral#1}}
\newcommand{\ignore}[1]{}

\def\ind#1#2{#1\cdots #2}
\usepackage{graphicx}

\begin{document}


\title{Dual Spaces of Resonance In Thick $p-$Branes}

\author{R. R. Landim$^1$, G. Alencar$^2$, M. O. Tahim$^2$, M. A. M. Gomes$^{3}$ and R. N. Costa Filho$^{1}$}

\affiliation{$^1$Departamento de F\'{\i}sica, Universidade Federal do Cear\'{a}, Caixa
Postal 6030, Campus do Pici, 60455-760, Fortaleza, Cear\'{a}, Brazil.\\$^2$ Universidade
Estadual do Cear\'a, Faculdade de Educa\c c\~ao, Ci\^encias e Letras do Sert\~ao Central, Quixad\'a,Cear\'a, Brazil.\\
$^3${Instituto Federal de Educa\c c\~ao, Ci\^encia e Tecnologia do Cear\'{a}-Av. Treze de Maio, 2081 - Benfica - Cep:60040-531 - Fortaleza - CE, Brazil.}
}

\date{\today}


\begin{abstract}

In this work we consider $q-$form fields in a $p-$brane embedded in a $D=(p+2)$
space-time. The membrane is generated by a domain wall in a Randall-Sundrum-like scenario.
We study conditions for localization of zero modes of these fields. The expression agrees
and generalizes the one found for the zero, one, two and three-forms in a $3-$brane. By a
generalization we mean that our expression is valid for any form in an arbitrary dimension
with codimension one. We also point out that, even without the dilaton
coupling, some form fields are localized in the membrane. The massive modes are considered
and the resonances are calculated using a numerical method. We
find that different spaces have identical resonance structures, which we call dual spaces
of resonances(DSR).

\end{abstract}


\maketitle

\section{Introduction}
Duality is a sort of symmetry that exists between two or more very special
theories. The remarkable fact about these kinds of symmetries is that when we make
the duality transformations over the important fields, we change the
regime of couplings in the theories involved. In other words, if we
have a theory whose coupling constant is $g$, it is possible to build
another theory with coupling constant $\frac{1}{g}$, i.e., we can do
strong coupling calculations in one model in perturbation approach if we change it by its dual.
The most basic example is the duality between the Sine-Gordon Model and
the Thirring Model. In this case, the solitons of the first model are
exchanged by the particles of the former \cite{Coleman:1974bu}. Another
examples are the network of dualities relating the five Superstring
theories and eleven-dimensional M-Theory \cite{Becker:2007zj}.
The most recent duality found, still in the Superstring context, is the
$AdS/CFT$ correspondence, where a $10$-dimensional Type $IIB$ superstring
in the $AdS_{5}\times S^{5}$ background is conjectured to be dual to a
$N=4$ Super-Yang Mills theory in $D=4$ \cite{Maldacena:1997re}. In this piece of work we
study tensorial fields in extra-dimensional spaces and find an
intriguing characteristic that is, in some sense, related to duality.
The fact is that we find evidences of an structure of resonances that have the same
characteristics but in different higher dimensional spaces with
different regimens of coupling constants.This last fact is the one
that made us call such a behavior as a "duality".

Antisymmetric tensor fields or simply differential forms arise naturally in string theory
\cite{Polchinski:1998rq,Polchinski:1998rr} and supergravity \cite{VanNieuwenhuizen:1981ae}
and play an important role in dualization \cite{Smailagic:1999qw,Smailagic:2000hr}. In
particular they appear in the $R-R$ sector of each of the type II string theories. These
tensor fields couple naturally to higher-dimensional extended objects, the $D-$branes and
are important for their stability. Besides this, they are related to the linking number of
higher dimensional knots \cite{Oda:1989tq}. The rank of these antisymmetric tensors is
defined by the dimension of the manifold\cite{Nakahara:2003nw}. Beyond this, these kind of
fields play an important role in the solution of the moduli stabilization problem of
string theory\cite{Kachru:2002he,Antoniadis:2004pp,Balasubramanian:2005zx}.

Because of these aspects, it is important to study higher rank tensor fields in membrane
backgrounds. In this direction antisymmetric tensor fields have already been considered in
models of extra dimension. Generally, the $q-$forms of highest rank do not have physical
relevance. This is due to the fact that when the rank increases, also increases the number
of gauge freedom. Such fact can be used to cancel the dynamics of the field in the
brane\cite{Alencar:2010vk}. The mass spectrum of the two and three-form have been studied,
for example, in Refs. \cite{Mukhopadhyaya:2004cc} and \cite{Mukhopadhyaya:2007jn} in a
context of five dimensions with codimension one. Later, the coupling between the two
and three-forms with the dilaton was studied, in different contexts, in
\cite{DeRisi:2007dn,Mukhopadhyaya:2009gp,Alencar:2010mi}.

In another direction soliton-like solutions are studied with increasing interest in
physics, not only in Condensed Matter, as in Particle Physics and Cosmology. In brane
models, they are used as mechanism of field localization, avoiding the appearance of the
troublesome infinities. Several kinds of defects in brane scenarios are considered in the
literature \cite{yves:a,yves:b,yves:c}. In these papers, the authors consider brane world
models where the brane is supported by a soliton solution to the baby Skyrme model or by
topological defects available in some models. As an example, in a recent paper, a model is
considered  for coupling fermions to brane and/or antibrane modeled by a kink antikink
system \cite{yves:d}. The localization of fields in a framework that consider the brane as
a kink has been studied for example in
\cite{Bazeia:2008zx,Bazeia:2007nd,Bazeia:2004yw,Bazeia:2003aw,Li:2010dy,Guo:2010az,Fu:2011pu,Liu:2011zy}. 
In this context the present authors have studied the issue of localization and resonances of a three-form
field in \cite{Alencar:2010hs} and in a separate paper  $q-$form fields in a scenario of a
$p-$brane with codimension two \cite{preparation}.

In these scenarios, some facts about localization of fields are known: The scalar
field($0-$form) is localizable, but the vector gauge fields ($1-$form), the Kalb-Ramond
field($2-$form) and the three-form field are not. The reason why this happens with the
vector field is that, in four dimensions, it is conformal and all information coming from
warp factors drops out, necessarily rendering a non-normalizable four dimensional
effective action. However, in the work of Kehagias and Tamvakis \cite{Kehagias:2000au}, it
is shown that the coupling between the dilaton and the vector gauge field produces
localization of the vector field. In analogy with the work of Kehagias and
Tamvakis\cite{Kehagias:2000au}, as cited above, the authors have also considered these
coupling with a three-form field\cite{Alencar:2010hs}, where a condition for localization
is found.

The facts above are even clear if we consider, in a general way, a $q-$form field in
an arbitrary dimension. This can be considered with and without the dilaton coupling. In
this reasoning, general conditions for localization can be found in an elegant
way. For example, for branes with space dimensions bigger than three, one should expect
that the gauge vector field is localizable even without the dilaton coupling. This is due
to the fact that this field is not conformal in these cases. In fact what we find is that,
bigger is the space-time dimension, bigger will be the order of the $q-$form localized
without the need of the dilaton. For the other cases the dilaton coupling is needed.

The main goal of this paper is to present the structure of dual resonance spaces.
However, standard calculations related to aspects of $q-$form fields living in the
background of a codimension one $p-$brane are necessary. Therefore, the paper is organized
as follows. Section 1 is devoted to discuss a solution of Einstein equation with
source given by a kink. In Section 2 we analyze how the gauge freedom can be used to
cancel the angular component of an arbitrary $q-$form. Yet in this section we discuss
aspects of localization of the fields involved. In the third section, we study  massive
modes, resonances and show aspects of dual spaces containing same resonance structures. At
the end, we discuss the conclusions and perspectives.

\section{The kink as a membrane}

We start our analysis studying the space-time background. Before analyzing the coupling
of the $q-$forms with the dilaton, it is necessary to obtain
a solution of the equations of motion for the gravitational field in the
background of the dilaton and the membrane. For such, we introduce the
following action, similar of the one used in \cite{Kehagias:2000au}:
\begin{equation}
S=\int d^{D}x \sqrt{-G}[2M^{3}R-\frac{1}{2}(\partial\phi)^{2}-\frac{1}{2}%
(\partial\pi)^{2}-V(\phi,\pi)],
\end{equation}
where $D=p+2$, $G$ is the metric determinant and $R$ is the Ricci scalar. Note that we are
working with a model containing two real scalar fields. The field $\phi$ plays the role of
membrane generator of the model while the field $\pi$ represents the dilaton. The
potential function depends on both scalar fields. It is assumed the following ansatz for
the spacetime metric:
\begin{equation}
ds^{2}=e^{2A_s(y)}\eta_{\mu\nu}dx^{\mu}dx^{\nu}+e^{2B_s(y)}dy^{2},
\end{equation}
where $\eta_{\mu\nu}=diag(-1,1,\cdots,1)$ is the metric of the $p$-brane, $y$ is the
codimension coordinate, and $s$ is a deformation parameter. The deformation parameter
controls the kind of deformed topological defect we want in order to mimic different
classes of membranes. The deformation method is based in deformations of the potential of
models containing solitons in order to produce new and unexpected solutions
\cite{Bazeia:2002xg}. As usual, capital Latin index represent the coordinates in the bulk
and Greek index, those on the $p$-brane.
The equations of motion are given by
\begin{equation}
\frac{1}{2}(\phi^{\prime})^{2}+\frac{1}{2}(\pi^{\prime})^{2}-e^{2B_s(y)}V(%
\phi,\pi)=24M^{3}(A_s^{\prime})^{2},
\end{equation}
\begin{equation}
\frac{1}{2}(\phi^{\prime})^{2}+\frac{1}{2}(\pi^{\prime})^{2}+e^{2B_s(y)}V(%
\phi,\pi)=-12M^{3}A_s^{\prime\prime}-24M^{3}(A_s^{\prime})^{2}+12M^{3}A_s^{%
\prime}B_s^{\prime},
\end{equation}
\begin{equation}
\phi^{\prime\prime}+(4A_s^{\prime}-B_s^{\prime})\phi^{\prime}=\partial_{\phi}V,
\end{equation}
and
\begin{equation}
\pi^{\prime\prime}+(4A_s^{\prime}-B_s^{\prime})\pi^{\prime}=\partial_{\pi}V.
\end{equation}

In order to solve this system, we use the so-called superpotential function $
W_s(\phi)$, defined by $\phi^{\prime}=\frac{\partial W_s}{\partial\phi}$,
following the approach of Kehagias and Tamvakis \cite{Kehagias:2000au}. The
particular solution regarded follows from choosing the potential $V_s(\phi,\pi)$ and superpotential $W_s(\phi)$ as
\begin{equation}
V_s=\exp{(\frac{\pi}{\sqrt{12M^{3}}})}\{\frac{1}{2}(\frac{\partial W_s}{%
\partial\phi})^{2}-\frac{5}{32M^{2}}W_s(\phi)^{2}\},
\end{equation}
and
\begin{equation}
W_s(\phi)=a\phi^2\left(\frac{s}{2s-1}(\frac{v}{\phi})^{1/s}-\frac{s}{2s+1}(\frac{\phi}{v})^{1/s}\right),
\end{equation}
where $a$ and $v$ are parameters to adjust the dimensionality. As pointed out in \cite{Kehagias:2000au}, this potential 
give us the desired soliton-like solution.In this way it is easy to
obtain first order differential equations whose solutions are solutions of the
equations of motion above, namely
\begin{equation}
\pi_s=-\sqrt{3M^{3}}A_s,
\end{equation}
\begin{equation}
B_s=\frac{A_s}{4}=-\frac{\pi_s}{4\sqrt{3M^{3}}},
\end{equation}
and
\begin{equation}
A_s^{\prime}=-\frac{W_s}{12M^{3}}.
\end{equation}
The solutions for these new set of equations are the following:

For $s=1$:
\begin{equation}
\phi(y)=v\tanh(ay),  \label{dilat}
\end{equation}
\begin{equation}
A(y)=-\frac{v^2}{72M^3}\left(4\ln\cosh(ay) +\tanh^2(ay)\right) \label{dilat2}
\end{equation}
and for $s>1$:
\begin{equation}
\phi(y)=v\tanh^s(ay/s),  \label{dilat1}
\end{equation}
\begin{eqnarray}
A_s(y)=&-&\frac{v^2}{12M^3}\frac{s}{2s+1}\tanh^{2s}\left(\frac{ay}{s}\right) \nonumber \\
&-& \frac{v^2}{6M^3}\left(\frac{s^2}{2s-1}-\frac{s^2}{2s+1}\right) \label{a} \\
&\times&\biggl{\{}\ln\biggl[\cosh\left(\frac{ay}{s}\right)\biggr]-
\sum_{n=1}^{s-1}\frac1{2n}\tanh^{2n}\left(\frac{ay}{s}\right)\biggr{\}}
\nonumber
\end{eqnarray}

An important observation to be made here is concerned to the appearance of a spacetime 
singularity due to the dilaton field. The Ricci scalar is given by
$$
R_s=-\left(8A_s^{\prime \prime }+18A_s^{\prime 2}\right)e^{\frac{\pi_s}{2\sqrt{3M^{3}}}}
$$
and the plot is given in the figure below
\begin{figure}[ht]
\centerline{\includegraphics[scale=1.0]{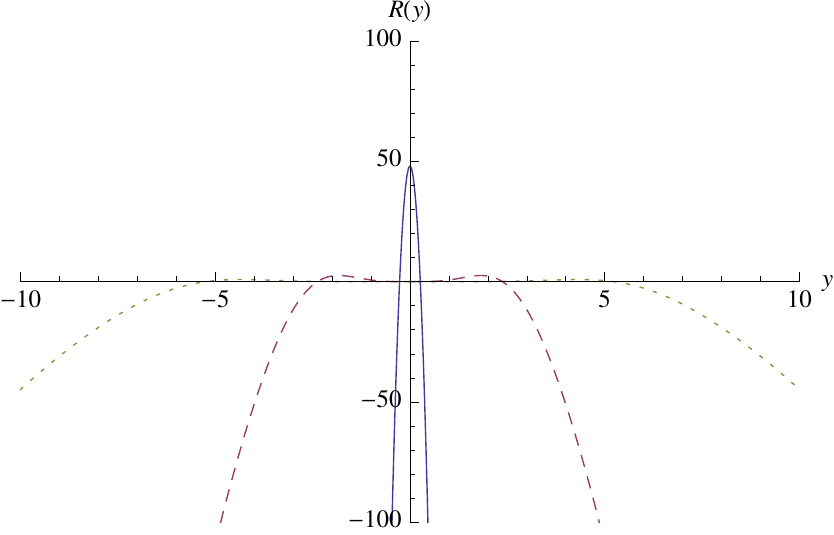}}
\caption{Ricci scalar for $s=1$ (lined) , $s=3$ (dashed) and $s=5$ (dotted) .}\label{fig}
\end{figure}

This is a naked singularity emerging at the infinity 
and in principle could jeopardize the issue of fields localization. However this singularity 
can be understood from a higher dimensional viewpoint. As an example,
a $D4$-brane in type II-A theory is singular because the presence of a diverging dilaton field. 
However, that singularity disappears when it is lifted to the eleventh dimensional supergravity viewpoint. 
Here, we changed our solution to six dimensions
$$
ds^2=e^{\frac{3A(y)}{2}} (-dt^2+dx_1^2+dx_2^2+dx_3^2+dz^2)+dy^2
$$
the dilaton turns out to be the radius of an extra $S^1$ direction (that is why it is called dilaton). 
The above higher dimensional metric is free of singularities. This interpretation also explain why, 
in this manuscript, we consider the same coupling between all  q-forms and the dilaton. This kind 
of solutions have been studied, for example, in \cite{Gherghetta:2000qi} and \cite{Oda:2000zc}. In fact, we used the string like solution of 
\cite{Oda:2000zc} to analyze the localization  q-forms in co-dimension two brane word \cite{Alencar:2010vk}. 
Although the objective of the present manuscript is not to deal with this singularity problem, this issue 
is widely discussed in the literature. It has been argued that the imposition of unitary boundary conditions 
renders these singularities harmless from the physical viewpoint \cite{HerreraAguilar:2009wc},\cite{Cohen:1999ia}, \cite{Gremm:1999pj},\cite{Gremm:2000dj}. 
Besides this aspect, there are reasons to believe that considering these kind of spaces as physically meaningful. 
There are proposals \cite{Gubser:1999vj,Giddings:2000mu} that five-dimensional
bulk gravity in the thin domain-wall case has an equivalent description in terms of a
cutoff four-dimensional conformal field theory on the domain-wall, on the lines of the
$AdS/CFT$ correspondence \cite{Maldacena:1997re}. From the viewpoint of $AdS/CFT$
correspondence, some singularities may have physical significance and it is very
interesting to see the role the dilaton field plays in this framework \cite{Gremm:2000dj}.

\section{Localization of $q-$form  Fields}

In this section we study $q-$form fields in the gravitational background of the last
section. We consider these fields with and without the dilaton coupling. It is important
to note that, when we exclude this coupling, the metric do not has the $B$ factor. With
these considerations, we must look for localization of forms in these frameworks. The
first important thing to be understood is the gauge symmetry of these fields. The fact  is
that, when the number of dimension increases, the number of gauge freedom also increases,
and this can be used to cancel the degrees of freedom in the visible brane. This analysis
was performed in details by the authors in a recent paper \cite{Alencar:2010vk}. Due the
importance of this for our present work, we give here a sketch of it. First, note that the
number of degrees of freedom (d.o.f.) of an antisymmetric tensor field with $q$ indices is
given by
$$
\left(\begin{array}{c}
D\\q
\end{array}\right)=\frac{D!}{q!\left(D-q\right)!}.$$

For the one form we have
$$
\delta X_{M}=\partial_{M}\phi$$
where $\phi$ is a scalar field. Therefore we have for the number of physical d.o.f.
\begin{eqnarray}
\left(\begin{array}{c}
D\\
1\end{array}\right)-\left(\begin{array}{c}
D\\
0\end{array}\right)=\left(\begin{array}{c}
D-1\\
1\end{array}\right). \label{one form}
\end{eqnarray}

In the above expression we have used the Stiefel relation. After this we claim that the number of physical d.o.f. of a $q-$form is given by
\begin{equation}
\left(\begin{array}{c}
D-1\\
q\end{array}\right). \label{dof}
\end{equation}
and proves that this is valid for a $(q+1)-$form. In this case the gauge
parameter will be an $q-$form and therefore, using the Stiefel relation,
we have for the physical d.o.f. of the $(q+1)-$form
\begin{eqnarray*}
&&\left(\begin{array}{c}
D\\
q+1\end{array}\right)-\left(\begin{array}{c}
D-1\\
q\end{array}\right)=\left(\begin{array}{c}
D-1\\
q+1\end{array}\right).
\end{eqnarray*}

The D-form has no dynamics. Using gauge symmetries, from the above result, we can see that the d.o.f. of the $(D-1)$ form can be made all null at the visible brane.
Therefore we must analyze only the cases $q=0,1,2,3...p$. The full action
for this field is given by
\begin{equation}
S_X=\int d^{D}x\sqrt{-G}[Y_{\ind{M_1}{M_{q+1}}}Y^{\ind{M_1}{M_{q+1}}}],
\end{equation}
where $Y_{M_1 \cdots M_{q+1}}=\partial _{\lbrack M_1}X_{M_2\dots M_{q+1}]}$ is the field
strength for the $q$-form $X$. We can use gauge freedom to fix $X_{\mu_1\cdots\mu_{q-1}
y}=\partial^{\mu_1 }X_{\ind{\mu_1}{\mu_q}}=0$ and we are left with the following two type
of terms
\begin{equation}
Y_{\ind{\mu_1}{\mu_{q+1}}}=\partial _{\lbrack \mu_1}X_{\ind{\mu_2}{\mu_{q+1}}]},
\end{equation}
\begin{equation}
Y_{y\ind{\mu_1}{\mu_{q}}}=\partial _{\lbrack y}X_{\ind{\mu_1}{\mu_q}}] =\partial_y X_{\ind{\mu_1}{\mu_q}}=X'_{\ind{\mu_1}{\mu_q}}.
\end{equation}

Using the above facts we obtain for the action
\begin{eqnarray*}
S_{X} = &\!\!\!\int d^{D}x\{[e^{((p-2q-1)A_s)}Y_{\mu_1\cdots\mu_{q+1}}
Y^{\mu_1 \cdots\mu_{q+1}}\\&+(q+1)e^{((p+1-2q)A_s)}{X'}_{\mu_1 \cdots\mu_{q}}%
{X'}^{\mu_1\cdots\mu_{q}}]\},
\end{eqnarray*}%
and the equation of motion are given by
\begin{eqnarray}
\partial _{\mu_1 }Y^{\mu_1 \cdots\mu_{q+1}}
+e^{((2q+1-p)A_s)}\partial _{y}[e^{((p+1-2q)A_s)}{X'}^{\mu_2\cdots\mu_{q+1}}]=0
\end{eqnarray}

We can now separate the $y$ dependence of the field using
\begin{equation}
X^{\ind{\mu_1}{\mu_q} }\left( x^{\alpha },y\right) =B^{\ind{\mu_1}{\mu_q}}\left(
x^{\alpha }\right) U\left( y\right).
\end{equation}

 Defining $Y^{\ind{\mu_1}{\mu_{q+1}}}=\tilde{Y}%
^{\ind{\mu_1}{\mu_{q+1}} }U$, where $\tilde{Y}$ stands for the four
dimensional field strength, we get
\begin{equation}
 \partial _{\mu_1 }\tilde{Y}^{\mu_1 \cdots\mu_{q+1}}+m^2B^{\ind{\mu_2}{\mu_{q+1}}}=0
\end{equation}
and
\begin{equation}
{U}''\left( y\right) -\left( (2q-p-1){A_s}'\right) {U}'\left(
y\right) =-m^{2}e^{-2A_s}U\left( y\right).
\end{equation}

In the above expression $m$ is a constant of separation of variable that represents the
mass of $q$-form $B$ in the $p$-brane. It is easy to see that $U=U_0$, with $U_0$ being a
constant, solves the above equation for $m=0$. In this particular case, the effective
action can be found easily to give us
\begin{equation}
S_{X}=\int dye^{{(p-2q-1)A_s}}U^{2}\int
d^{p}x[\tilde{Y}_{\alpha \mu \lambda \gamma }\tilde{Y}^{\alpha \mu \nu
\lambda }].
\end{equation}

Using now our solution for $A_s$ we have that the integration in the extra dimension is
finite for $q<(p-1)/2$. It becomes clear now that only the scalar field is localized for
$p=3$. We also see that for $p=4$, for example, the vector field is localized. As
commented in the introduction, this was expected because this field is not conformal in
this space dimension. Despite of this, the dilaton is yet needed for the localization of
higher order forms. The coupling is inspired in string theory and we have for the action
\begin{equation}
S_X=\int d^{D}x\sqrt{-G}e^{-\lambda \pi_s }[Y_{\ind{M_1}{M_{q+1}}}Y^{\ind{M_1}{M_{q+1}}}],
\end{equation}
with the same definitions used above. Repeating the same steps as before we obtain
\begin{eqnarray*}
S_{X} =&&\!\!\!\!\!\!\!\!\!\!\int d^{D}x\{[e^{((p-2q-1)A_s+B_s-\lambda\pi_s)}Y_{\mu_1\cdots\mu_{q+1}}
Y^{\mu_1 \cdots\mu_{q+1}}\\+&&\!\!\!\!\!\!\!(q+1)e^{((p+1-2q)A_s-B_s-\lambda\pi_s)}{X'}_{\mu_1 \cdots\mu_{q}}
{X'}^{\mu_1\cdots\mu_{q}}]\}.
\end{eqnarray*}

The equation of motion takes the form
\begin{eqnarray}
\partial _{\mu_1 }Y^{\mu_1 \cdots\mu_{q+1}}
+e^{((2q+1-p)A_s-B_s+\lambda\pi_s)}\partial _{y}[e^{((p+1-2q)A_s-B_s-\lambda\pi_s)}{X'}^{\mu_2\cdots\mu_{q+1}}]=0,
\end{eqnarray}
and performing the same separation of variable we get
\begin{equation}
{U}''\left( y\right) -\left( (2q-p-1){A_s}'+{B_s}'+\lambda\pi_s'\right) {U}'\left(
y\right) =-m^{2}e^{2\left( B_s-A_s\right) }U\left( y\right) . \label{Udilaton}
\end{equation}

Again we can see that $U=U_0$, with $U_0$ being a constant, solves the above equation for
$m=0$. The effective action can again be found easily and is given by
\begin{equation}
S_{X}=\int dye^{\left({(p-2q-1)A_s+B_s-\lambda\pi_s}\right)}U^{2}\int
d^{p}x[\tilde{Y}_{\alpha \mu \lambda \gamma }\tilde{Y}^{\alpha \mu \nu
\lambda }].
\end{equation}

Using now our solution for $A_s$, $B_s$ and $\pi_s$ we have that, when
$\lambda >(8q-4p+3)/4\sqrt{3m^3}$, the integration in the extra dimension is finite for
all $q$-forms in any $p$-brane. It is important to stress that for $p=3$, the above
expression reproduces the same conditions found for the one, two and three form found in
the literature.

\section{The Massive Modes}

For the massive modes, the best way to make the analysis is to transform the equation
(\ref{Udilaton}) in a Schr\"odinger type equation. It is easy to see that an equation of
the form
\begin{equation}
 \left(\frac{d^2}{dy^2}-P_s'(y)\frac{d}{dy}\right)U(y)=-m^2Q_s(y)U(y),\label{pq}
\end{equation}
can be put in a Schr\"odinger form
\begin{equation}
 \left(-\frac{d^2}{dz^2}+V_s(z)\right)\bar{U}(z)=m^2\bar{U}(z),\label{schlike}
\end{equation}
through the transformations
\begin{equation}
 \frac{dz}{dy}=f_s(y), \quad U(y)=\Omega_s(y)\bar{U}(z),
\end{equation}
with
\begin{equation}
 f_s(y)=\sqrt{Q_s(y)}, \quad \Omega_s(y)=\exp(P_s(y)/2)Q_s(y)^{-1/4},
\end{equation}
and
\begin{equation}
 V_s(z)=\left(P_s'(y)\Omega_s'(y)-\Omega_s''(y)\right)/\Omega_s f_s^2
\end{equation}
where the prime is a derivative with respect to $y$. Now using the equation (\ref{Udilaton}) we obtain
\begin{eqnarray}
 &&V_s(z)=e^{3A_s/2}\left((\frac{\alpha^2}{4}-\frac{9}{64})A_s'(y)^2-(\frac{\alpha}{2}+\frac{3}{8})A_s''(y)\right),  \nonumber \\ && f_s(y)=e^{-\frac{3A_s}{4}},\quad \Omega(y)=e^{(\frac{\alpha}{2}+\frac{3}{8})A_s},
\end{eqnarray}
where $\alpha=(8q-4p-3)/4-\lambda\sqrt{3M^3}$, and we take $y$ of function of $z$ through $z(y)=\int_0^yf(\eta)d\eta$.
We show bellow the graphic of $V_s(z)$ by considering  $v^2/{72M^3}=1$ and $a=1$ for various $\alpha$ values.

\begin{figure}[ht]
\centerline{\includegraphics[scale=0.8]{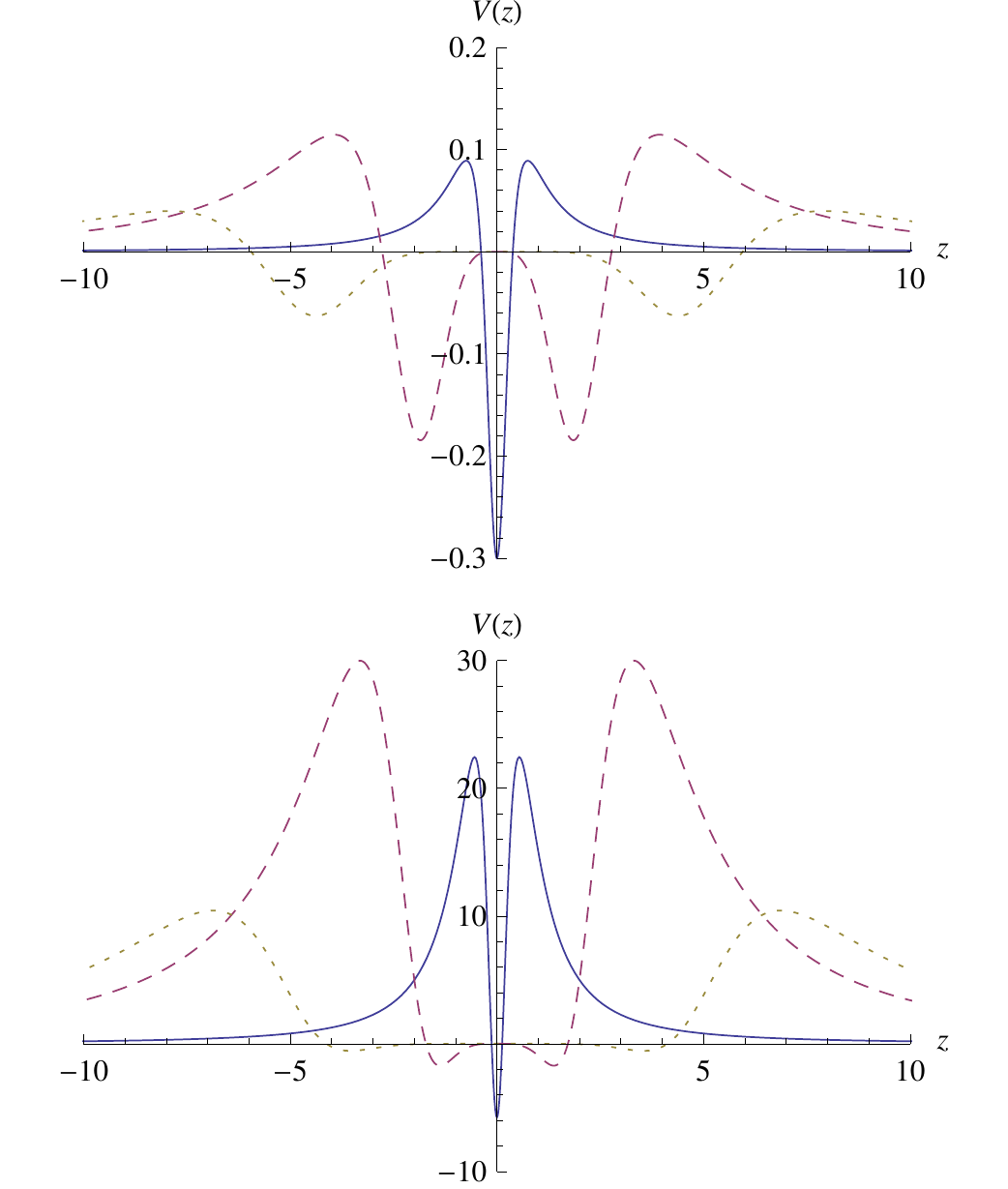}}
\caption{Potential of the Schroedinger like equation for $s=1$ (lined) scaled by 1/10, $s=3$ (dashed), $s=5$ (dotted) for  $\alpha=-1.75$ (top)
and $\alpha=-20$ (bottom).}\label{fig1}
\end{figure}
In order to analyze resonances, we must compute transmission coefficients($T$)
numerically, which gives a clear physical interpretation about what happens to a free wave
that interact with the membrane. The idea of the existence of a resonant mode is that for
a given mass the potential barrier is transparent to the particle, i.e., the transmission
coefficient has a peak at this mass value. That means that the amplitude of the
wave-function has a maximum value at $z=0$ and the probability to find this KK mode inside
the membrane is higher. In order to obtain these results numerically, we have developed a
computational code to compute transmission coefficients for the above potential profile. A
more extensive and detailed analysis of resonances with transmission coefficients will be
given in a separate paper by the authors \cite{preparation}.

\begin{figure}[ht]
\centerline{\includegraphics[scale=0.7]{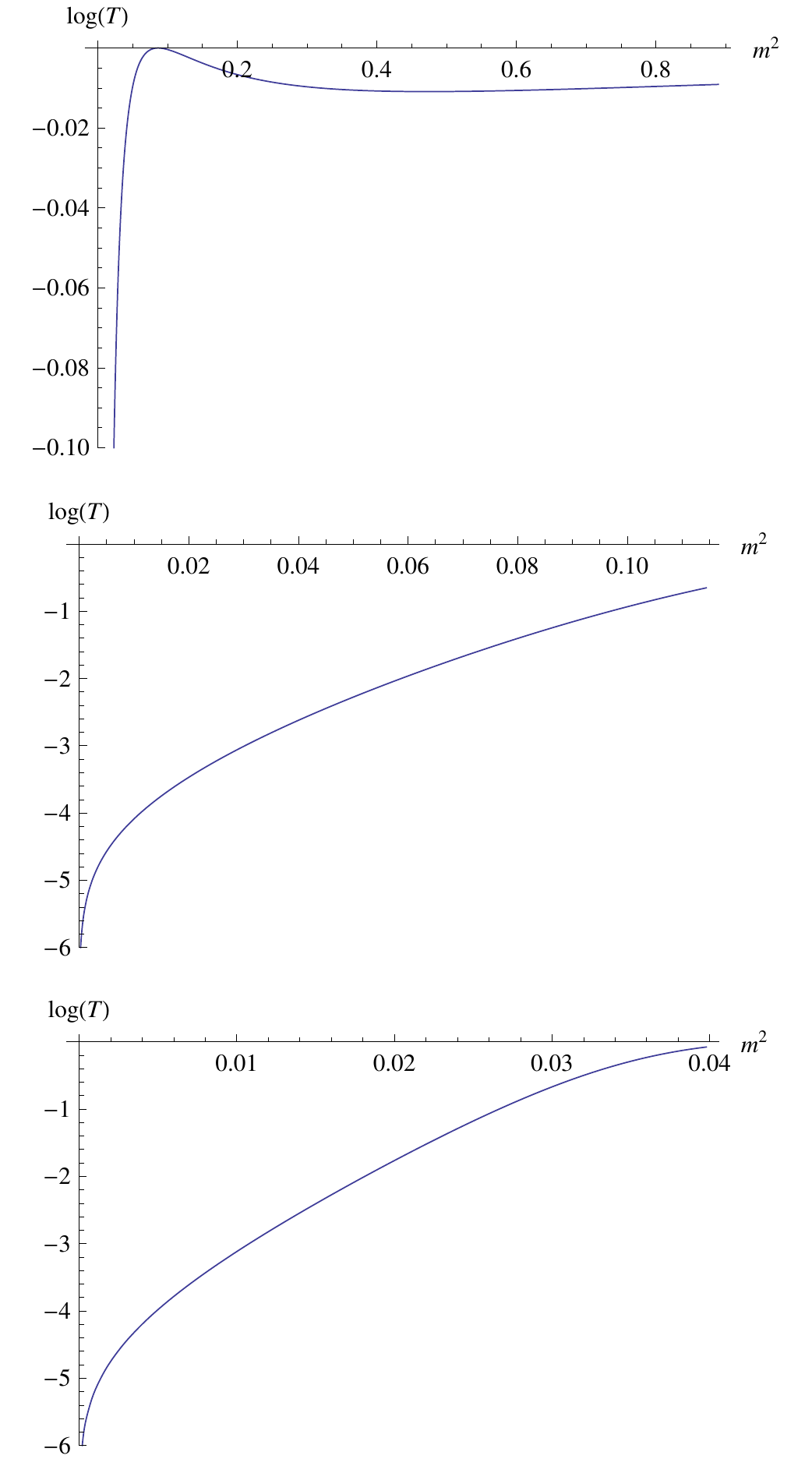}}
\caption{Logarithm of the transmission coefficient for $s=1$ (top), $s=3$ (middle), $s=5$ (bottom) for  $\alpha=-1.75$ .}\label{fig2}
\end{figure}

\begin{figure}[ht]
\centerline{\includegraphics[scale=0.7]{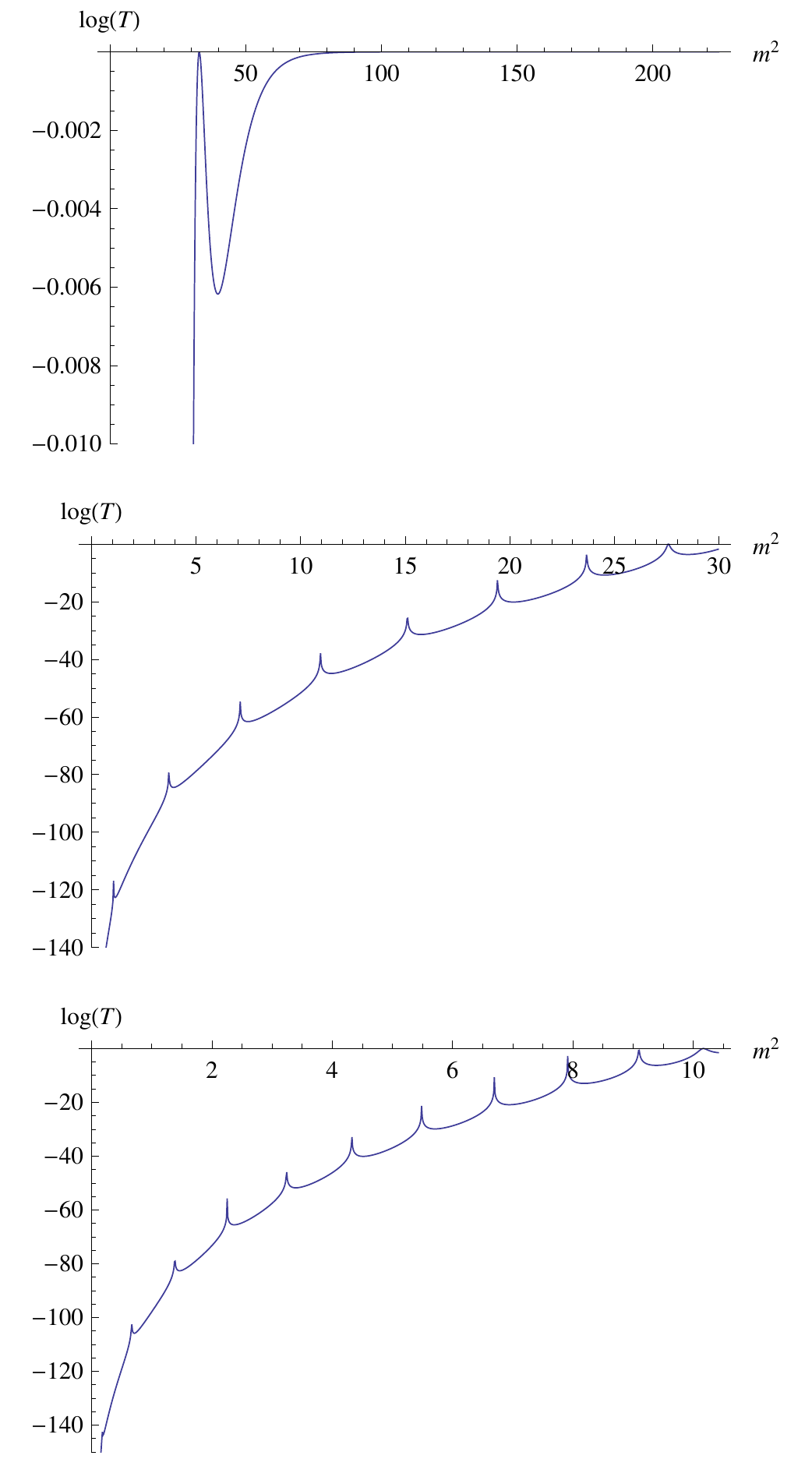}}
\caption{Logarithm of the transmission coefficient for $s=1$ (top), $s=3$ (middle), $s=5$ (bottom) for  $\alpha=-20$ }\label{fig3}
\end{figure}

In FIG. \ref{fig2} we give the plots of $\log(T)$ for $\alpha=-1.75,$ and considering the case $s=1,3,5$
As mentioned before, we find resonances for this value of $\alpha$ only for $s=1$.
In FIG. \ref{fig3} we show the plot for $\alpha=-20$

and we can see how this alter the existence of resonances. For this values we have an interesting structure of resonances, witch is similar to that found
in quantum mechanical problems. As we stress in our separate paper, the existence of resonances depends strongly of the shape of the potential.
In our case, therefore, the resonance is driven by the choice of the parameter $\alpha$.

\subsection{The case $p=3$}

The case for $p=3$ is doubtless the more important one and several issues deserve a careful analysis. First, the generalization
employed here simplifies the problem and clearly shows how the condition $\lambda >(8q-4p+3)/4\sqrt{3m^3}$, necessary for localization of the zero mode, 
changes for different $p$ and $q$. For comparison, previous articles found this condition for $q=1,2,3$ separately \cite{Cruz:2010zz}\cite{Cruz:2009ne}\cite{Alencar:2010hs}. 
Therefore, we gain a considerable simplification and understanding for all those cases with our generalization. 
Moreover, we found resonances that was never observed before in \cite{Cruz:2010zz} and \cite{Cruz:2009ne}. This is mainly due the fact that we realized 
that the resonances depends strongly of the parameter $\alpha$. In \cite{Cruz:2010zz} the authors study a one form and found only one resonance peak. 
By changing the parameter we discovered much more resonant modes FIG. \ref{fig2}, indicating that the higher the parameter more resonances we get. 
The same study was performed by the same authors in ref. \cite{Cruz:2009ne}, but for the two form. They founded the localization condition for the zero mode, 
that agrees with our expression.  After this, they studied the possibility of resonances and found just one resonance peak. Again, with our 
generalization, besides the peak found by them we found a lot of other peaks when we change the parameter $\alpha$ as in FIG. \ref{fig3}. The case $q=3$ was 
presented in a recent paper by the present authors\cite{Alencar:2010hs}, where the rich structure of resonances was first found. It is important to note 
here that all analysis pointed  above can be visualized in the two graphics given below FIG. \ref{fig2} and FIG. \ref{fig3} . 
This was only possible due the general expression for $\alpha$ that we are presenting here. 

\subsection{Dual Spaces of Resonance}
As a byproduct the present authors has observed that for a fixed value of $\alpha$ and $\lambda$, a $q$-form in a $p$-brane
has the same resonance structure of a $q'$-from in a $p'$-brane if  $2q-p=2q'-p'$. All
graphics above are plotted for $2q-p=3$ and are the same as that with a three-form in
3-brane \cite{Alencar:2010hs}. Also for a fixed $\lambda$ and $q$ the resonance peaks
increase for larger values of the dimension of the brane. There are an interesting
situation with $\alpha$ constant. We can have models with small values of $\lambda$
displaying  the same resonance structure. We see in table bellow some values of $p$, $q$
and $\lambda\sqrt{3M^3}$ with  $\alpha=-20$.

\begin{table}[h]
\begin{center}
\begin{tabular}{|c|c|}
\hline
p & q\\ \hline
3 & 3\\ \hline
9 & 6\\ \hline
21 & 12 \\ \hline
\end{tabular}
\hspace{2cm}
\begin{tabular}{|c|c|}
\hline
p & q\\ \hline
20 & 1\\ \hline
22 & 2\\ \hline
24 & 3 \\ \hline
\end{tabular}
\caption{In the left $q=(p+3)/2$, $\lambda\sqrt{3M^3}=89/4$ and in the rigth $q=(p-18)/2$, $\lambda\sqrt{3M^3}=5/4$.}
\end{center}
\end{table}

As we can see, we have dual spaces of resonance(DSR): they have the same resonance
structure. It is interesting to note that these DSR admits situations with large and low
coupling constant. This is similar to the $S$ duality of string theory. The difference is
that here we have an infinite class of spaces, with a certain $\lambda$, which has the
same resonant mass peaks as another infinite class, with a different $\lambda$. It is not
clear to the authors if there is some structure or symmetry behind these DSR that could
explain why they appear. We let this question for a future work.

\section{Conclusions and Perspectives}

In this paper we have studied the issue of localization of $q-$form fields in a $p-$brane
embedded in a $D=(p+2)$ space-time. The membrane was described by a kink, a codimension
one topological defect, and related topological objects called deformed defects.
Furthermore, we have considered a gravitational background where the dilaton field plays
an important role. We have obtained the localization of the zero modes of several
antisymmetric fields regarding specific peculiarities for each dimension. For example, for
branes with space dimensions bigger than three, we found that the gauge vector field is
localizable even without the dilaton coupling. This is due to the fact that this field is
not conformal in these cases. For other types of forms, the dilaton coupling is still
needed. We have calculated massive states through the related quantum mechanical problem
and have analyzed the appearance of possible resonances. In this case we have made use of
numerical computations of transmission coefficients through the membrane. The existence of
resonances depends strongly of the shape of the potential, as is well known.  We have been
able to find several peaks of resonant modes for specific parameters that models the
potential and for specific dimensionality of the spacetime. In particular, as an important
and intriguing result, we have found a class of resonances that have the same behavior,
but living in spacetimes with different dimensions and with different dilaton coupling
constant. Because of this last fact, we call this set of resonances in this model as dual
spaces of resonances(DSR). However, it is not at all understood how this appears from the
usual way to treat dualities in field theories. We let these sort of discussions for a
future work.

\vspace{1cm}

\section*{Acknowledgment}
We would like to thank: Prof. Francisco Pimentel for usefull discussions, and the Laboratório de
\'oleos pesados for the hospitality. We also acknowledge the financial
support provided by Funda\c c\~ao
Cearense de Apoio ao Desenvolvimento Cient\'\i fico e Tecnol\'ogico
(FUNCAP), the Conselho Nacional de Desenvolvimento Cient\'\i fico e Tecnol\'ogico (CNPq) and FUNCAP/CNPq/PRONEX.

This paper is dedicated to the memory of my wife  Isa\-bel Mara (R. R. Landim)


\begin{thebibliography}{99}


\bibitem{Coleman:1974bu}
  S.~R.~Coleman,
  Phys.\ Rev.\  D {\bf 11}, 2088 (1975).

\bibitem{Becker:2007zj}
  K.~Becker, M.~Becker and J.~H.~Schwarz,
{\it  Cambridge, UK: Cambridge Univ. Pr. (2007) 739 p}


\bibitem{Maldacena:1997re}
  J.~M.~Maldacena,
  Adv.\ Theor.\ Math.\ Phys.\  {\bf 2} (1998) 231
  [Int.\ J.\ Theor.\ Phys.\  {\bf 38} (1999) 1113]
  [arXiv:hep-th/9711200].

\bibitem{Polchinski:1998rq}
  J.~Polchinski,
``String theory. Vol. 1: An introduction to the bosonic string,''
{SPIRES
entry} {\it  Cambridge, UK: Univ. Pr. (1998) 402 p}

\bibitem{Polchinski:1998rr}
  J.~Polchinski,
``String theory. Vol. 2: Superstring theory and beyond,''
{SPIRES entry} {\it  Cambridge, UK: Univ. Pr. (1998) 531 p}

\bibitem{VanNieuwenhuizen:1981ae}
  P.~Van Nieuwenhuizen,
  Phys.\ Rept.\  {\bf 68}, 189 (1981).

\bibitem{Smailagic:1999qw}
  A.~Smailagic and E.~Spallucci,
  Phys.\ Rev.\  D {\bf 61}, 067701 (2000)
  [arXiv:hep-th/9911089].

\bibitem{Smailagic:2000hr}
  A.~Smailagic and E.~Spallucci,
  Phys.\ Lett.\  B {\bf 489}, 435 (2000)
  [arXiv:hep-th/0008094].

\bibitem{Oda:1989tq}
  I.~Oda and S.~Yahikozawa,
  ``Linking Numbers And Variational Method,''
  Phys.\ Lett.\  B {\bf 238}, 272 (1990).

\bibitem{Nakahara:2003nw}
  M.~Nakahara,
  ``Geometry, topology and physics,''
{\it  Boca Raton, USA: Taylor} \& {\it Francis (2003) 573 p}

\bibitem{Kachru:2002he}
  S.~Kachru, M.~B.~Schulz and S.~Trivedi,
  ``Moduli stabilization from fluxes in a simple IIB orientifold,''
  JHEP {\bf 0310}, 007 (2003)
  [arXiv:hep-th/0201028].

\bibitem{Antoniadis:2004pp}
  I.~Antoniadis and T.~Maillard,
  ``Moduli stabilization from magnetic fluxes in type I string theory,''
  Nucl.\ Phys.\  B {\bf 716}, 3 (2005)
  [arXiv:hep-th/0412008].

\bibitem{Balasubramanian:2005zx}
  V.~Balasubramanian, P.~Berglund, J.~P.~Conlon and F.~Quevedo,
  ``Systematics of Moduli Stabilisation in Calabi-Yau Flux Compactifications,''
  JHEP {\bf 0503}, 007 (2005)
  [arXiv:hep-th/0502058].
\bibitem{Gubser:1999vj}
  S.~S.~Gubser,
  Phys.\ Rev.\  D {\bf 63} (2001) 084017
  [arXiv:hep-th/9912001].
  
\bibitem{Giddings:2000mu}
  S.~B.~Giddings, E.~Katz and L.~Randall,
  JHEP {\bf 0003}, 023 (2000)
  [arXiv:hep-th/0002091].
  
\bibitem{Gremm:2000dj}
  M.~Gremm,
  Phys.\ Rev.\  D {\bf 62}, 044017 (2000)
  [arXiv:hep-th/0002040].

\bibitem{Gherghetta:2000qi}
  T.~Gherghetta and M.~E.~Shaposhnikov,
  Phys.\ Rev.\ Lett.\  {\bf 85}, 240 (2000)
  [arXiv:hep-th/0004014].

\bibitem{Alencar:2010vk}
  G.~Alencar, R.~R.~Landim, M.~O.~Tahim and K.~C.~Mendes,
  ``Antisymmetric Tensor Fields in Codimension Two Brane-World,''
  arXiv:1009.1183 [hep-th].

\bibitem{HerreraAguilar:2009wc}
  A.~Herrera-Aguilar, D.~Malagon-Morejon, R.~R.~Mora-Luna and U.~Nucamendi,
  Mod.\ Phys.\ Lett.\  A {\bf 25}, 2089 (2010)
  [arXiv:0910.0363 [hep-th]].

\bibitem{Cohen:1999ia}
  A.~G.~Cohen and D.~B.~Kaplan,
  Phys.\ Lett.\  B {\bf 470}, 52 (1999)
  [arXiv:hep-th/9910132].

\bibitem{Gremm:1999pj}
  M.~Gremm,
  Phys.\ Lett.\  B {\bf 478}, 434 (2000)
  [arXiv:hep-th/9912060].



\bibitem{Mukhopadhyaya:2004cc}
  B.~Mukhopadhyaya, S.~Sen, S.~Sen and S.~SenGupta,
  ``Bulk Kalb-Ramond field in Randall Sundrum scenario,''
  Phys.\ Rev.\  D {\bf 70}, 066009 (2004)
  [arXiv:hep-th/0403098]

\bibitem{Mukhopadhyaya:2007jn}
  B.~Mukhopadhyaya, S.~Sen and S.~SenGupta,
  ``Bulk antisymmetric tensor fields in a Randall-Sundrum model,''
  Phys.\ Rev.\  D {\bf 76}, 121501 (2007)
  [arXiv:0709.3428 [hep-th]].


\bibitem{DeRisi:2007dn}
  G.~De Risi,
  ``Bouncing cosmology from Kalb-Ramond Braneworld,''
  Phys.\ Rev.\  D {\bf 77}, 044030 (2008)
  [arXiv:0711.3781 [hep-th]].

\bibitem{Mukhopadhyaya:2009gp}
  B.~Mukhopadhyaya, S.~Sen and S.~SenGupta,
  ``A Randall-Sundrum scenario with bulk dilaton and torsion,''
  Phys.\ Rev.\  D {\bf 79}, 124029 (2009)
  [arXiv:0903.0722 [hep-th]].



\bibitem{Alencar:2010mi}
  G.~Alencar, M.~O.~Tahim, R.~R.~Landim {\it et al.},
  Phys.\ Rev.\  {\bf D82}, 104053 (2010).
  [arXiv:1005.1691 [hep-th]].


\bibitem{yves:a}
Y.~Brihaye, T. ~Delsate, N. ~Sawado and Y.~Kodama,
     [arXiv:hep-th/1007.0736]


\bibitem{yves:b}
Y.~Brihaye, T. ~Delsate,
 Class.Quant.Grav {\bf 24},1279-1292 (2007)
   [arXiv:gr-qc/0605039]

\bibitem{yves:c}
Y.~Brihaye, T. ~Delsate, B. ~Hartman
 Phys.Rev.D {\bf 74},044015(2006)
   [arXiv:hep-th/0602172]


\bibitem{yves:d}
Y.~Brihaye and T.~Delsate,
 Phys.Rev.D{\bf 78}, 025014 (2008)
  [arXiv:hep-th/0803.1458]

\bibitem{Bazeia:2008zx}
  D.~Bazeia, A.~R.~Gomes, L.~Losano and R.~Menezes,
  ``Braneworld Models of Scalar Fields with Generalized Dynamics,''
  Phys.\ Lett.\  B {\bf 671}, 402 (2009)
  [arXiv:0808.1815 [hep-th]].

\bibitem{Bazeia:2007nd}
  D.~Bazeia, A.~R.~Gomes and L.~Losano,
  ``Gravity localization on thick branes: a numerical approach,''
  Int.\ J.\ Mod.\ Phys.\  A {\bf 24}, 1135 (2009)
  [arXiv:0708.3530 [hep-th]].

\bibitem{Bazeia:2004yw}
  D.~Bazeia, F.~A.~Brito and A.~R.~Gomes,
  ``Locally localized gravity and geometric transitions,''
  JHEP {\bf 0411}, 070 (2004)
  [arXiv:hep-th/0411088].

\bibitem{Bazeia:2003aw}
  D.~Bazeia, C.~Furtado and A.~R.~Gomes,
  ``Brane structure from scalar field in warped spacetime,''
  JCAP {\bf 0402}, 002 (2004)
  [arXiv:hep-th/0308034].

\bibitem{Li:2010dy}
  H.~-T.~Li, Y.~-X.~Liu, Z.~-H.~Zhao {\it et al.},
  Phys.\ Rev.\  {\bf D83}, 045006 (2011).
  [arXiv:1006.4240 [hep-th]].

\bibitem{Guo:2010az}
  H.~Guo, Y.~-X.~Liu, S.~-W.~Wei {\it et al.},
  [arXiv:1008.3686 [hep-th]].

\bibitem{Fu:2011pu}
  C.~-EFu, Y.~-X.~Liu, H.~Guo,
  
  [arXiv:1101.0336 [hep-th]].

\bibitem{Liu:2011zy}
  Y.~-X.~Liu, H.~Guo, C.~-EFu {\it et al.},
  
  [arXiv:1101.4145 [hep-th]].

\bibitem{Alencar:2010hs}
  G.~Alencar, R.~R.~Landim, M.~O.~Tahim {\it et al.},
  Phys.\ Lett.\  {\bf B693}, 503-508 (2010).
  [arXiv:1008.0678 [hep-th]].

\bibitem{preparation}
  G.~Alencar, R.~R.~Landim, M.~O.~Tahim, R.~N.~Costa Filho and M~A.~M.~Gomes,
  In preparation

\bibitem{Oda:2000zc}
  I.~Oda,
  ``Localization of matters on a string-like defect,''
  Phys.\ Lett.\  B {\bf 496}, 113 (2000)
  [arXiv:hep-th/0006203].


\bibitem{Oda:2000zj}
  I.~Oda,
  ``Bosonic fields in the string-like defect model,''
  Phys.\ Rev.\  D {\bf 62}, 126009 (2000)
  [arXiv:hep-th/0008012].


\bibitem{Olasagasti:2000gx}
  I.~Olasagasti and A.~Vilenkin,
  ``Gravity of higher-dimensional global defects,''
  Phys.\ Rev.\  D {\bf 62}, 044014 (2000)
  [arXiv:hep-th/0003300].

\bibitem{Kehagias:2000au}
  A.~Kehagias and K.~Tamvakis,
  ``Localized gravitons, gauge bosons and chiral fermions in smooth spaces
  generated by a bounce,''
  Phys.\ Lett.\  B {\bf 504}, 38 (2001)
  [arXiv:hep-th/0010112].
\bibitem{Bazeia:2002xg}
  D.~Bazeia, L.~Losano and J.~M.~C.~Malbouisson,
  ``Deformed defects,''
  Phys.\ Rev.\  D {\bf 66}, 101701 (2002)
  [arXiv:hep-th/0209027].

\bibitem{Cruz:2010zz}
  W.~T.~Cruz, M.~O.~Tahim, C.~A.~S.~Almeida,
  Phys.\ Lett.\  {\bf B686}, 259-263 (2010).

\bibitem{Cruz:2009ne}
  W.~T.~Cruz, M.~O.~Tahim, C.~A.~S.~Almeida,
  Europhys.\ Lett.\  {\bf 88}, 41001 (2009).
  [arXiv:0912.1029 [hep-th]].



\end{thebibliography}
\end{document}